\documentclass{jfm}

\usepackage{graphicx}
\usepackage{natbib}
\usepackage{amsmath,amssymb}

\ifCUPmtlplainloaded \else
  \checkfont{eurm10}
  \iffontfound
    \IfFileExists{upmath.sty}
      {\typeout{^^JFound AMS Euler Roman fonts on the system,
                   using the 'upmath' package.^^J}%
       \usepackage{upmath}}
      {\typeout{^^JFound AMS Euler Roman fonts on the system, but you
                   dont seem to have the}%
       \typeout{'upmath' package installed. JFM.cls can take advantage
                 of these fonts,^^Jif you use 'upmath' package.^^J}%
      }
  \else
  \fi
\fi

\ifCUPmtlplainloaded \else
  \checkfont{msam10}
  \iffontfound
    \IfFileExists{amssymb.sty}
      {\typeout{^^JFound AMS Symbol fonts on the system, using the
                'amssymb' package.^^J}%
       \usepackage{amssymb}%
         \let\leq=\leqslant
         
      }{}
  \fi
\fi

\ifCUPmtlplainloaded \else
  \IfFileExists{amsbsy.sty}
    {\typeout{^^JFound the 'amsbsy' package on the system, using it.^^J}%
     \usepackage{amsbsy}}
    {\providecommand\boldsymbol[1]{\mbox{\boldmath $##1$}}}
\fi

\newcommand\Pran{\mbox{\textit{Pr}}} 

\newcommand\Nu{\mbox{\textit{Nu}}}   
\newcommand\Ra{\mbox{\textit{Ra}}}   
\newcommand\Ro{\mbox{\textit{Ro}}}   
\newcommand\Ek{\mbox{\textit{Ek}}}   
\newcommand\Ta{\mbox{\textit{Ta}}}   
\newcommand\Fr{\mbox{\textit{Fr}}}   

\newcommand\rms{{r\!m\!s}}   
\newcommand\kin{{k\!i\!n}}   
\newcommand\pol{{p\!o\hspace*{-0.5pt}l}}   
\newcommand\tor{{t\!o\hspace*{-0.5pt}r}}   
\newcommand\crit{{c\hspace*{-0.5pt}r}}   

\newsavebox{\astrutbox}
\sbox{\astrutbox}{\rule[-5pt]{0pt}{20pt}}

\usepackage{graphicx}
\usepackage{overpic}
\renewcommand{\vec}{\boldsymbol}
\usepackage{sistyle}
\title{Toroidal and poloidal energy in rotating  Rayleigh--B\'enard~convection}

\author[S. Horn and O. Shishkina]%
{Susanne Horn$^{1,2}$%
  \thanks{Email address for correspondence: susanne.horn@ds.mpg.de}\ns
and Olga Shishkina$^{1,2}$}

\affiliation{}

\affiliation{$^1$Max Planck Institute for Dynamics and Self-Organization, Am Fa\ss berg 17, 37077 G\"ottingen, Germany\\[\affilskip]
$^2$Institute of Aerodynamics and Flow Technology, German~Aerospace~Center~(DLR), Bunsenstra\ss e 10, 37073 G\"ottingen, Germany}

\pubyear{2010}
\volume{650}
\pagerange{119--126}
\date{?; revised ?; accepted ?. - To be entered by editorial office}

\usepackage{xcolor}
\definecolor{pink}{RGB}{208,32,144}

\begin{document}

\maketitle

\begin{abstract}
We consider rotating Rayleigh--B\'enard convection of a fluid with a Prandtl number of $\Pran = 0.8$  in a cylindrical cell with an aspect ratio $\Gamma = 1/2$. Direct numerical simulations were performed for the Rayleigh number range $10^5 \leq \Ra \leq 10^9$ and the inverse Rossby number range $0 \leq 1/\Ro \leq 20$. We propose a method to capture regime transitions based on the decomposition of the velocity field into toroidal and poloidal parts. We identify four different regimes. First, a buoyancy dominated regime occurring as long as the toroidal energy $e_{\tor}$ is not affected by rotation and remains equal to that in the non-rotating case, $e^0_{\tor}$. Second, a rotation influenced regime, starting at rotation rates where $e_{\tor} > e^0_{\tor}$ and ending at a critical inverse Rossby number $1/\Ro_{\crit}$ that is determined by the balance of the toroidal and poloidal energy, $e_{\tor} = e_{\pol}$. Third, a rotation dominated regime, where the toroidal energy $e_{\tor}$ is larger than both, $e_{\pol}$ and $e^0_{\tor}$.
Fourth, a geostrophic turbulence regime for high rotation rates where the toroidal energy drops below the value of non-rotating convection.
\end{abstract}

\begin{keywords}
\end{keywords}

\section{Introduction}
Turbulent flows, driven by thermal convection and  affected by rotation, are ubiquitous phenomena in geo- and astrophysics. Examples are the convection in stars, in the interior of gaseous planets and in the Earth's atmosphere and oceans, to mention only a few. And even though, these phenomena are also shaped by other processes, such as magnetic fields, stratification or liquid-gas phase transition, convection under the influence of the Coriolis force is crucial to their description. Thus to increase our fundamental understanding of the physics behind it, we study rotating Rayleigh--B\'enard convection, i.e. a fluid rotated about its vertical axis which is heated from below and cooled from above.\par
The commonly used control parameters of rotating Rayleigh--B\'enard convection are the Rayleigh number $\Ra$, the Prandtl number $\Pran$ and the convective Rossby number $\Ro$, defined by
\begin{equation} 
\Ra = \frac{\alpha g H^3 \Delta}{\kappa \nu},\quad \Pran = \frac{\nu}{\kappa}, \quad \Ro = \frac{\sqrt{\alpha g\Delta H}}{2 \Omega H},
\end{equation} 
 where $\alpha$ is the isobaric expansion coefficient, $g$ the acceleration due to gravity, $H$ the vertical distance between the top and bottom plate, $\Delta$ the imposed adverse temperature difference, $\kappa$ the thermal diffusivity, $\nu$ the viscosity and $\Omega$ the angular speed.  Instead of $\Ro$ occasionally also the Taylor number $\Ta$ and the Ekman number $\Ek$ are  used to characterise the importance of rotation, which are given by
 \begin{eqnarray}
\Ta &=&  \left(\frac{2 \Omega H^2}{\nu}\right)^2 = \frac{\Ra}{\Pran \Ro^2},\\
\Ek &=&  \frac{\nu}{2 \Omega H^2} = \Ro \Pran^{1/2} \Ra^{-1/2} = \Ta^{-1/2}.
 \end{eqnarray}
Apart from that, the geometry of the container, in particular its aspect ratio, also plays an important role. However, the preferred aspect ratio has changed over the years. Starting from investigating convection in cylindrical containers with large diameter-to-height aspect ratios, $\Gamma = D/H$, recent developments in numerical and experimental studies, rather go to smaller and smaller $\Gamma$ of $0.5$ \cite[]{He2012,Ahlers2012} or even $0.23$ \cite[]{Stevens2011}. Most of the earlier studies were about the onset of convection and pattern formation \cite[]{Chandrasekhar1961}, thus, the aim was to mimic an infinite lateral extent, where analytical relations are available. On the contrary, most of the current investigations focus on turbulent thermal convection, including the transition to the so-called ``ultimate state'' \citep{Grossmann2011}, thus, the aim is to achieve high $\Ra$ and, hence, practical considerations demand a small $\Gamma$. The development to smaller $\Gamma$ is not only true for ``ordinary'', but also for rotating convection \cite[]{Oresta2007,Stevens2012,Ecke2013}. Yet, the finite size has serious implications for rotating Rayleigh--B\'enard convection. Not only, does the destabilising effect of the lateral wall yield a lower critical $\Ra$ for the onset of convection at fast rotation rates \citep{Buell1983}  because of drifting wall modes \citep{Zhong1991, Ecke1992, Kuo1993, Herrmann1993, Goldstein1993, Goldstein1994}, but $\Gamma$ also determines the bifurcation point $\Ro_b$, at which, for $\Pran \gtrsim 1$ and higher $\Ra$, heat transfer enhancement sets in \cite[]{Weiss2010,Weiss2011a}.\par
The increased heat transport, expressed in terms of the Nusselt number $\Nu$, is usually used as an indicator for the different turbulent states occurring in rotating turbulent thermal convection, suggesting a division into three regimes \cite[]{Kunnen2011}. In the weak rotation regime, $\Nu$ remains nearly constant, but as soon as $1/\Ro$ is increased to values above $1/\Ro_b$, after a sharp onset, a continuous increase of $\Nu$ is observed for moderate rotation rotates, which coincides with the generation of columnar vortex structures \cite[]{Weiss2010,Stevens2011}. After it has reached a peak, which marks the transition to the regime of strong rotation, it drops rapidly with the rotation rate due to the suppression of vertical velocity fluctuations \cite[cf. also the recent review by][]{Stevens2013a}. However, this classification of regimes is only valid for fluids with $\Pran \gtrsim 1$; for $\Pran \lesssim 1$ no heat transfer enhancement is expected \cite[]{Stevens2010b}.\par 
As the change of $\Nu$ is closely connected to the columnar vortices, the number of vortices serves as another criterion to determine the point where rotation dominates over buoyancy. However, extracting these vortices is relatively cumbersome, and involves a certain arbitrariness in choosing what constitutes a vortex. Furthermore, for $\Pran < 1$ the larger diffusivity results in only short vortices that dissipate quickly when  they reach the bulk, which complicates matters. Conversely, one can also look at the large-scale circulation (LSC) or more specifically at the rotation rate when it breaks-down \citep{Kunnen2008a, Weiss2011b, Stevens2012}. In experiments, this is frequently obtained by analysing the temperature signal at the sidewall. However, one has to be careful with two-vortex states or multiple-roll states occurring in Rayleigh--B\'enard cells with small aspect ratios. Evidently, also the crossover of the boundary layer thicknesses \citep{Rossby1969, King2009, King2012} cannot be applied to fluids with $\Pran < 1$, where the thermal boundary layer is thicker than the viscous one even without rotation.\par
Here, we offer an alternative  method for the characterisation of the different regimes in rotating Rayleigh--B\'enard convection. Motivated by the work by \cite{Breuer2004}, who have shown, that the toroidal and poloidal energy are characteristic for the distinctive types of dynamics in low and high Prandtl number flows in non-rotating convection, namely, that the toroidal energy is highest for fluids with $\Pran \lesssim 1$ \cite[]{Breuer2004}, and vanishes for $\Pran = \infty$ \cite[]{Busse1967b}, we analyse the contribution of the toroidal and poloidal energy in rotating convection.  This is a very natural approach. The poloidal energy is the energy contained in cellular or roll motion, such as the LSC or double-roll states, i.e. the dominant motion without rotation. The toroidal energy, on the other hand, is contained in swirling motion in the horizontal plane, i.e. with a vertical vorticity \citep{Olson1991}, which is the dominant motion in rotating convection. This means, we are able to distinguish different regimes of rotating convection based on global quantities, namely the time and volume averaged toroidal and poloidal energy without a restriction to certain Prandtl numbers or aspect ratios.
\section{Numerical method}
We study rotating Rayleigh--B\'enard convection by means of direct numerical simulations (DNS) using a fourth order finite volume code for cylindrical domains. Details about the code can be found in \cite{Shishkina2005} and \cite{Horn2013a}. Additionally, we implemented a term describing the Coriolis force, whereas the centrifugal potential can be incorporated in the reduced pressure, and, hence, does not need to be considered explicitly.\par
 We neglect any effects due to centrifugal buoyancy whose importance can be estimated by calculating the Froude number
\begin{equation}
 \Fr = \frac{\Omega^2 D}{2g} = \frac{\Ra \Gamma}{8 \Ro^2} \frac{\kappa \nu}{g H^3}.
\end{equation}
Since numerically dimensionless equations are solved, we use the parameters of the High-Pressure Convection Facility (HPCF) at the Max Planck Institute for Dynamics and Self-Organization in G\"ottingen, Germany, to evaluate $\Fr$. It is a cylindrical cell, with a height of $H = \SI{2.24}{m}$ and a diameter of $D = \SI{1.12}{m}$, i.e. $\Gamma = 1/2$, filled with sulfur hexafluoride (SF$_{\mbox{6}}$) at pressures between \SI{2}{bar} and \SI{19}{bar} \citep{He2012, Ahlers2012}. At a pressure of \SI{2}{bar}, the viscosity of SF$_{\mbox{6}}$ is given by $\nu = \SI{1.2 \times {10^{-6}}}{m^2 s}$ and its thermal diffusivity by $\kappa = \SI{1.6 \times {10^{-6}}}{m^2 s}$, i.e. $\Pran \approx 0.8$. The gravitational acceleration in G\"ottingen is approximately \SI{9.81}{m s^{-2}}. Thus, for the highest $\Ra$ and $1/\Ro$, namely $10^9$ and $20$, the Froude number is as low as $\Fr = 4.4 \times 10^{-4}$. Hence, since $\Fr \ll 1$ it is justifiable to set $\Fr \equiv 0$ \citep{Zhong2009}. However, it should be noted, that the Rayleigh numbers in the HPFC are typically much larger than the ones we can attain in our DNS.\par
The governing equations of the problem are the continuity equation for incompressible flows, the Navier--Stokes equations in Oberbeck--Boussinesq approximation and the energy equation,
\begin{eqnarray}
\vec{\nabla} \cdot \tilde{\vec{u}} &=& 0, \\
D_{\tilde{t}} \tilde{\vec{u}} &=& - \tilde{\rho}^{-1} \vec{\nabla} \tilde{p}  + \tilde{\nu} \vec{\nabla}^2 \tilde{\vec{u}}  + 2 \tilde{\vec{\Omega}} \times \tilde{\vec{u}} + \tilde{g} \tilde{\alpha} \tilde{T} \hat{\vec{e}}_z,\\
D_{\tilde{t}} \tilde{T} &=& \tilde{\kappa} \vec{\nabla}^2 \tilde{T},
\end{eqnarray}%
where $D_t$ denotes the substantial derivative, $\hat{\vec{e}}_z$ the unit vector in vertical direction, $\vec{u}$ the velocity, $T$ the temperature, $p$ the reduced pressure, $\vec{\Omega}~=~\Omega \hat{\vec{e}}_z$ the angular velocity and $\rho$ the density.  The tilde denotes that the quantity is given in its dimensional representation. However, numerically we solve dimensionless equations in cylindrical coordinates $(r,\,\phi,\,z)$. All variables are replaced by their scaled counterparts, using the radius $\hat{R}$, the buoyancy velocity $(\hat{g} \hat{\alpha} \hat{R} \hat{\Delta})^{1/2}$, the temperature difference $\hat{\Delta}$ and the material properties at the mean temperature as reference scales. The reference time is then given by $\hat{R}/(\hat{g} \hat{\alpha} \hat{R} \hat{\Delta})^{1/2}$ and the reference pressure by $\hat{\rho} \hat{g} \hat{\alpha} \hat{R} \hat{\Delta}$. In vector notation, the dimensionless governing equations read
\begin{eqnarray}
\vec{\nabla} \cdot {\vec{u}} &=& 0, \label{eq:NS1}\\
D_{{t}} {\vec{u}} &=& - \vec{\nabla} {p}  + \Ra^{-1/2} \Pran^{1/2} \gamma^{-3/2} \vec{\nabla}^2 {\vec{u}}  + \Ro^{-1} \gamma^{1/2} {\hat{\vec{e}}_z} \times {\vec{u}} + {T} {\hat{\vec{e}}}_z, \label{eq:NS2}\\
D_{{t}} {T} &=&   \Ra^{-1/2} \Pran^{-1/2} \gamma^{-3/2} \vec{\nabla}^2 {T},\label{eq:NS3}
\end{eqnarray}%
with the radius-to-height aspect ratio $\gamma$. The sidewall is assumed to be adiabatic and the top and bottom plates are isothermal with $T\hspace*{-3pt}\mid_{z=H} = T_t = -0.5$ and $T\hspace*{-3pt}\mid_{z = 0} = T_b= 0.5$. The boundary conditions for the velocity are no-slip at all walls.\par
All simulation were performed for a fluid with $\Pran = 0.8$, corresponding to SF$_6$ or air, in a slender cylinder with $\Gamma = 1/2$ for $\Ra \in [10^5,10^9]$ and rotation rates $1/\Ro \in [0,20]$.
The resolution is chosen to fulfil the requirements by \citet{Shishkina2010}. The meshes are non-equidistant in radial and vertical direction, with a clusterisation of grid points close to the walls. To guarantee enough points in the boundary layers, the grid points were denser clustered for smaller $\Ro$, as it is well-known, that the viscous boundary layer becomes thinner as the rotation rate increases \cite[e.g.][]{Kunnen2008a, Stevens2010a}.  This means, that in the Ekman type viscous boundary layer in the rotating case the same criterion for the number of grid points was applied as in the non-rotating case. The details of all simulation parameters and the numerical resolution can be found in table~\ref{tab:parameters}.
\begin{table}
  \begin{center}
\def~{\hphantom{0}}
  \begin{tabular}{p{1.cm}p{1.cm}p{1.cm}p{6.5cm}p{2cm}}
    $\Pran$  & $~\Gamma$ & $\Ra$  &  \hfil $1/\Ro$ & $N_r \times N_\phi \times N_z$ \\[3pt]
       $0.8$   & $1/2$ & $10^5$ &  \parbox{6.5cm}{$\{0.0,0.07,0.14,0.23,0.47,0.71,1.41,2.36,\\~3.33,4.71,5.89,7.07,8.84,10.1,14.14\}$} & $~11 \times~ 32 \times ~34$\\
       $0.8$   & $1/2$ & $10^6$ &  \parbox{6.5cm}{$\{0.0,0.07,0.1,0.14,0.2,0.35,0.47,0.67, 1.0, \\~1.41,1.67,2.0,2.13,2.5,2.83,3.33,3.54,4.0, \\~4.71,5.0,6.67,7.07,8.33,10.0, 14.29,20.0\}$} & $~17 \times ~64 \times ~68$\\
       $0.8$   & $1/2$ & $10^7$ &  \parbox{6.5cm}{$\{0.0,0.07,0.14,0.71,1.41,2.0,2.83,4.71,\\~7.07,14.14\}$} & $~33 \times128 \times132$\\
       $0.8$   & $1/2$ & $10^8$ &  \parbox{6.5cm}{$\{0.0,0.07,0.14,0.2,0.71,1.0,1.41,2.0,2.36,\\~3.33,5.0,7.07,10.0,14.14,20.0\}$} & $~80 \times256 \times320$\\
       $0.8$   & $1/2$ & $10^9$ &  \parbox{6.5cm}{$\{0.0,0.07,1.41,2.0,2.36,2.83,4.0,7.07,\\~10.0,14.14,20.0\}$} & $192 \times 512 \times 768$\\
  \end{tabular}
  \caption{Parameters used in the DNS presented here, including the computational mesh size $N_r \times N_\phi \times N_z$.}
  \label{tab:parameters}
  \end{center}
\end{table}
\section{Decomposition of the velocity field}
 In our DNS the Navier--Stokes equations \eqref{eq:NS1}--\eqref{eq:NS3} are solved in primitive variables, hence in order to obtain the toroidal and poloidal energy, we analyse the instantaneous velocity fields every half dimensionless time unit. That means, we decompose the solenoidal velocity field $\vec{u}$ into a poloidal field $\vec{\pi}$ and a toroidal field $\vec{\tau}$ with the defining scalars $\xi$ and $\psi$, respectively, \cite[]{Chandrasekhar1961,Breuer2004}
\begin{equation}
 \vec{u} = \vec{\pi} + \vec{\tau} = \vec{\nabla} \times \vec{\nabla} \times (\xi \hat{\vec{e}}_z) +  \vec{\nabla} \times (\psi \hat{\vec{e}}_z). \label{eq:dec}
\end{equation}
This decomposition is also called Mie decomposition or Mie representation of the vector field $\vec{u}$ \citep{Backus1986}. In cylindrical component notation, equation \eqref{eq:dec} reads
\begin{align}
u_r &= \pi_r \hspace{1pt} + \tau_r\hspace{1pt} = \partial_r \partial_z \xi + \textstyle \frac{1}{r} \partial_\phi \psi \label{eq:dec1},\\
u_\phi &= \pi_\phi + \tau_\phi = \textstyle \frac{1}{r} \partial_\phi \partial_z \xi  - \partial_r \psi \label{eq:dec2},\\
u_z &= \pi_z \hspace{1pt} + \tau_z\hspace{1pt} =  \textstyle - \frac{1}{r} \partial_r (r \partial_r \xi) - \frac{1}{r^2} \partial_\phi \partial_\phi \xi. \label{eq:dec3}
\end{align}
The equations \eqref{eq:dec1} and \eqref{eq:dec2} can be combined, and expressed in terms of the vertical component of the vorticity $\vec{\omega} = \vec{\nabla} \times \vec{u}$,
\begin{equation}
\omega_z = - \textstyle \frac{1}{r} \left(\partial_r \left(r u_\phi\right) - \partial_\phi u_r  \right) =  \textstyle - \frac{1}{r} \partial_r (r \partial_r \psi) - \frac{1}{r^2} \partial_\phi \partial_\phi \psi \label{eq:om}.
\end{equation}
The operator $ \textstyle - \frac{1}{r} \partial_r (r \partial_r ) - \frac{1}{r^2} \partial_\phi \partial_\phi \equiv \Delta_{r\phi}$ is the horizontal Laplacian. The equations \eqref{eq:dec3} and \eqref{eq:om} are thus two-dimensional Poisson equations
\begin{eqnarray}
 \Delta_{r\phi} \xi + u_z &=& 0, \label{eq:poiss1}\\
 \Delta_{r\phi} \psi + \omega_z &=& 0. \label{eq:poiss2}
\end{eqnarray}
Therefore, these scalars are also called velocity potentials, in analogy to, e.g., electrodynamics. However, the scalars $\xi$ and $\psi$ are not uniquely defined. In fact, to the toroidal potential $\psi$ an arbitrary horizontal harmonic function, i.e. any solution of the corresponding Laplace equation, can be added. The poloidal potential $\xi$ is only determined up to an arbitrary function of $z$ \cite[see e.g.][]{Marques1993}. Thus, there is a gauge freedom for the boundary conditions. The most simple and commonly used gauge condition for $\xi$ is
\begin{equation}
\left.\xi\right|_{r=R}=0\label{eq:bc1}
\end{equation}
\citep{Marques1993, Boronski2007}. This gauge in combination with (\ref{eq:dec1}) and the no-slip condition on the velocity, $\left.u_r\right|_{r=R}=0$, yields
\begin{equation}
\partial_\phi \psi =0. 
\end{equation}
Hence, $\psi$ needs to be constant along the contour $r =R$ for a constant $z$ and we can also set consistently
\begin{equation}
\left.\psi\right|_{r=R}=0. \label{eq:bc2}
\end{equation}
Furthermore, we have
\begin{equation}
\left.\xi\right|_{z=0} = \left.\xi\right|_{z=H} = 0 \mbox{ and } \left.\psi\right|_{z=0} = \left.\psi\right|_{z=H} = 0  
\end{equation}
\citep{Marques1993}. This choice of gauge is the most convenient one, because it implies Dirichlet boundary conditions on $r = R$ on the two equations \eqref{eq:poiss1} and \eqref{eq:poiss2}. Hence, the problem of solving these two Poisson equations becomes well-posed, and the unique solutions for the poloidal and toroidal scalar field are given by
\begin{align}
\xi(r,\phi,z) &= \int\limits_0^{2\pi} \!\! \int\limits_0^Ru_z(\zeta,\eta,z) G(r,\phi,\zeta,\eta) \zeta d \zeta d \eta, \label{eq:xi} \\
\psi(r,\phi,z) &= \int\limits_0^{2\pi} \!\! \int\limits_0^R \omega_z(\zeta,\eta,z) G(r,\phi,\zeta,\eta) \zeta d \zeta d \eta, \label{eq:psi}
\end{align}
where $\zeta$ and $\eta$ are integration variables and $G(r,\phi,\zeta,\eta)$ is the Green's function
\begin{equation*}
G(r,\phi,\zeta,\eta)  = \frac{1}{4\pi} \ln\left(\frac{r^2 \zeta^2 - 2 R^2 r \zeta \cos(\phi-\eta) + R^4}{R^2 (r^2 - 2r \zeta \cos(\phi -\eta) + \zeta^2)}\right). 
\end{equation*}
 Albeit the fact, that this solution is analytically exact, the large mesh size of the numerically obtained flow fields makes the solving computationally expensive and for higher $\Ra$, i.e. $\Ra \gtrsim 10^8$, infeasible. Numerically, it is more efficient to solve the Poisson equations directly.
This was done by adapting the well-tested \texttt{fishpack90} \cite[]{Swarztrauber1975} solver to double precision and our non-equidistant meshes. The solver is based on the generalized Buneman algorithm. Special care is also required at the cylinder axis where in cylindrical coordinates one always faces the problem of the mathematical, but not physical singularity, caused by terms involving $1/r$. But this can be elegantly overcome by calculating \eqref{eq:xi} and \eqref{eq:psi} at $r = 0$  directly, utilizing that the Green's function in this case simplifies to
\begin{equation}
G(0,\phi,\zeta,\eta)  = \frac{1}{2\pi} \ln\left(\frac{R}{\zeta}\right).
\end{equation}
Thus, by prescribing the analytical solution at $r = 0$ as numerical boundary condition, a smooth scalar field is guaranteed. It is worth noting, that this does not impose any additional physical boundary or gauge condition, 
but is a direct consequence of the Dirichlet boundary conditions \eqref{eq:bc1} and \eqref{eq:bc2} and is, hence, merely a numerical trick.
Eventually, we are able to calculate the total kinetic energy $e_{\kin}$, the poloidal energy $e_{\pol}$ and the toroidal energy $e_{\tor}$, defined by
\begin{eqnarray}
e_{\kin}&=& \langle u_r^2 + u_\phi^2 + u_z^2 \rangle_{V,t}, \label{eq:ekin}\\
e_{\pol} &=& \langle \pi_r^2 + \pi_\phi^2 + \pi_z^2 \rangle_{V,t}, \label{eq:epol}\\ 
e_{\tor} &=& \langle \tau_r^2 + \tau_\phi^2 + \tau_z^2 \rangle_{V,t}, \label{eq:etor}
\end{eqnarray}%
where $\langle \cdot \rangle_{V,t}$ denotes averaging in time $t$ and over the whole volume $V$.
\section{Identifying transitions in SF{$_{\mbox{6}}$} in a $\boldsymbol{\Gamma = 1/2}$ cell}
We propose, that the toroidal and poloidal energy can be used to universally capture transitions in rotating Rayleigh--B\'enard convection. We present how this method can be applied to thermal convection of a fluid with $\Pran = 0.8$, corresponding to SF{$_{\mbox{6}}$ that is confined in a slender cylindrical cell with $\Gamma = 1/2$.
\subsection{Nusselt number and characteristic flow properties}
Most of the recent experiments and numerical simulations on rotating Rayleigh--B\'enard convection were conducted in water with $3 \lesssim \Pran \lesssim 7$ \cite[]{King2009,Zhong2010,Weiss2011b,Kunnen2010a,Stevens2009a}. One of the reasons for this might be, that only for fluids with $\Pran > 1$ columnar vortex structures, sometimes called Ekman vortices \cite[]{Stevens2010b,Weiss2010}, occur that extend from one horizontal wall to the other \cite[]{Horn2011b}. Via Ekman pumping, the vortices are able to significantly enhance the heat transport compared to the non-rotating case, where the heat flux is usually expressed in terms of the Nusselt number, 
\begin{equation}
 \Nu = \left(\Ra \Pran \gamma \right)^{1/2} \langle u_z T \rangle - \gamma^{-1} \left\langle \partial_z T \right\rangle \label{eq:Nu}.
\end{equation}
But for fluids with $\Pran < 1$, these vortices are much shorter, and do not form a regular grid. The reason is, that the thermal diffusivity is larger than the kinematic viscosity, thus, heat can spread in the bulk, making Ekman pumping less effective \cite[]{Stevens2010b}.
\citet{Stevens2010b} found no heat transfer enhancement at all for $\Pran = 0.7$, $\Ra = 10^8$, $0.1 \leq 1/\Ro \leq 10.0$ and  $\Gamma = 1.0$, however, \citet{Oresta2007} found a slightly higher $\Nu$ for very similar simulation parameters, $\Pran = 0.7$, $\Ra = 2 \times 10^8$,  $0.1 \leq 1/\Ro \leq 33.3$ but $\Gamma = 0.5$, which they attributed to Ekman pumping. Figure \ref{fig:Nu}(a) shows the  Nusselt number for the rotating case normalized by the one in the non-rotating case, $\Nu/\Nu^0$, for our DNS  for $10^5 \leq \Ra \leq 10^9$ and $0.07 \leq 1/\Ro \leq 20.0$. In addition, experimental results by \cite{Ecke2013} for $\Ra =6.2 \times 10^9$ in a cylindrical convection cell with $\Gamma = 0.5$ and helium with $\Pran = 0.7$ are shown for comparison. The Nusselt number $\Nu$ is also presented in figure \ref{fig:Nu}(b), but as function of the Taylor number.\par
For $\Ra = 10^5$, we have steady convection for all Rossby numbers considered, except for $1/\Ro = 14.1$, where convection is completely suppressed due to rotation and heat is transported by conduction alone. For $\Ra = 10^6$, convection is chaotic and unsteady for $1/\Ro \lesssim 1.67$, for  $1.67 \gtrsim 1/\Ro \gtrsim 2.5$ we found oscillatory convection and for even faster rotation rates, i.e. $1/\Ro \gtrsim 2.5$, convection is steady. For $\Ra = 10^7$, we have turbulent convection for low rotation rates, but again, however, for even faster rotation rates, i.e. $1/\Ro = 14.1$ we observed steady convection. Finally, for $\Ra = 10^8$ and $\Ra = 10^9$ our applied rotation was never rapid enough to completely suppress turbulent fluctuations.\par
The general behaviour of $\Nu$ with increasing rotation rate is very similar for all $\Ra$, i.e. it is almost constant for slow rotation and then drops rapidly at a certain rotation rate. There is also a very slightly increased $\Nu$ for $1/\Ro \lesssim 2$, which is due to the stabilizing effect of rotation, which suppresses reversals and changes from a one-roll state to a double-roll state, that occur more frequently for $\Gamma = 0.5$ than for $\Gamma = 1.0$. For $\Ra = 10^6$ there is more switching between these different states than for the other $\Ra$ which is a possible explanation for the evident deviation of the behaviour of the Nusselt number for that particular Rayleigh number. It was also found by \cite{Oresta2007}, in numerical simulations with parameters very similar to ours at $\Ra = 9 \times 10^5$ and $\Pran = 0.7$. The continuous decrease of $\Nu$ with increasingly high rotation rate is expected from the Taylor--Proudman theorem \cite[]{Taylor1921,Proudman1916}. It predicts the suppression of flow variations along the axis of rotation. Although strictly speaking it is not designated to the highly non-linear and time-dependent case of rotating Rayleigh--B\'enard convection, the reduced heat transport can be understood with it. In figure \ref{fig:Nu}(a) also the prediction by \citet{Weiss2010} and \citet{Weiss2011a} based on a phenomenological Ginzburg--Landau model is shown. They have shown, that for fluids with $\Pran \gtrsim 1$, at 
\begin{equation}
 \frac{1}{\Ro_b} = \frac{a}{\Gamma} \left(1 +\frac{b}{\Gamma}\right), \; a = 0.381,\, b = 0.061
\end{equation}
a bifurcation corresponding to the onset of Ekman vortex formation and Nusselt number enhancement occurs. This bifurcation is a finite-size effect, and gives  $1/\Ro_b = 0.86$ for $\Gamma = 0.5$. For $\Pran = 0.8$  and $\Ra = 10^5$ and $\Ra = 10^7$ the Nusselt number starts to decrease at the point $1/\Ro_b$, but now columnar vortices were observed. However, for $\Ra = 10^8$ and $\Ra = 10^9$, we found short columnar vortices and the Nusselt number also appears to increase slightly and then drops at a critical inverse Rossby number $2.36 \lesssim 1/\Ro_{\crit}  \lesssim 3.33$. When $\Nu$ is plotted against $\Ta$ as in figure~\ref{fig:Nu}~(b), the vertex of the curve, marked with filled symbols, seems to match all $\Ra$. In figure~\ref{fig:Nu}~(a) the same rotation rate is indicated by a grey shaded area. The transition at $1/\Ro_{\crit}$ is consistent with the one found empirically by \cite{Ecke2013} at $1/\Ro_1 \approx  2.86$. Furthermore, \cite{Ecke2013} suggested that after this initial decrease a more rapid decrease occurs after $1/\Ro_2 \approx 8.33$. An accurate identification of $\Ro_{\crit}$ solely based on the Nusselt number is nonetheless difficult.\par
\begin{figure}
\centering
\begin{overpic}{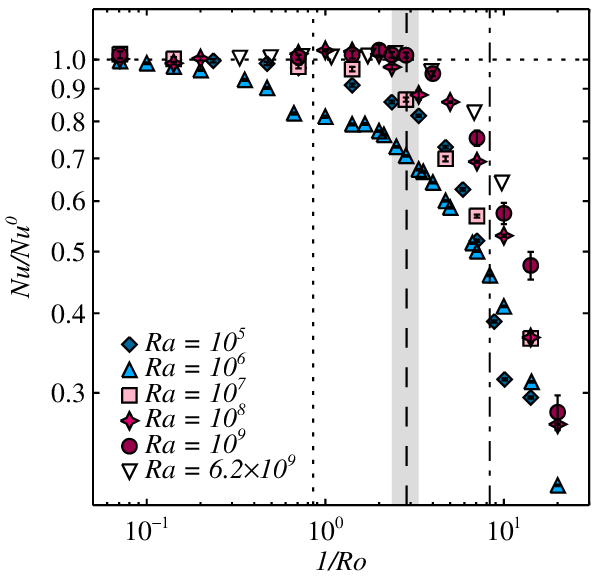}
 \put(0,92){(a)}
\end{overpic}
\hfill
\begin{overpic}{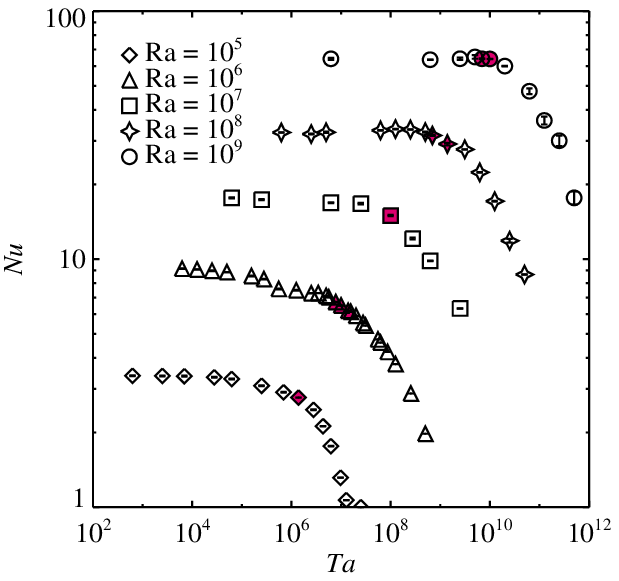}
  \put(0,92){(b)}
\end{overpic}
\caption{\label{fig:Nu}(Colour online) (a) Nusselt number for the rotating case normalized by the one in the non-rotating case $\Nu/\Nu^0$ as function of the inverse Rossby number $1/\Ro$ for $\Ra \in \{10^5, 10^6, 10^7, 10^8, 10^9\}$ obtained by DNS; experimental data by \cite{Ecke2013} for $\Ra = 6.2 \times 10^9$ and $\Pran = 0.7$ are shown for comparison. The vertical dotted line shows the prediction by \citet{Weiss2010,Weiss2011a}, $1/\Ro_b = 0.86$, the vertical long-dashed and dashed-dotted line mark the proposed transition by \cite{Ecke2013} at $1/\Ro_1 \approx 2.83$ and $1/\Ro_2 \approx 8.33$, respectively. The grey shaded area shows where $e_{\pol} \approx e_{\tor}$ at $2.36 \lesssim 1/\Ro_{\crit}  \lesssim 3.33$. (b) Nusselt number $\Nu$ as function of the Taylor number $\Ta$. The filled (pink) symbols show where $e_{\pol} \approx e_{\tor}$ and are the same as marked by the grey shaded area in figure (a).}
\end{figure}
\begin{figure}
\includegraphics{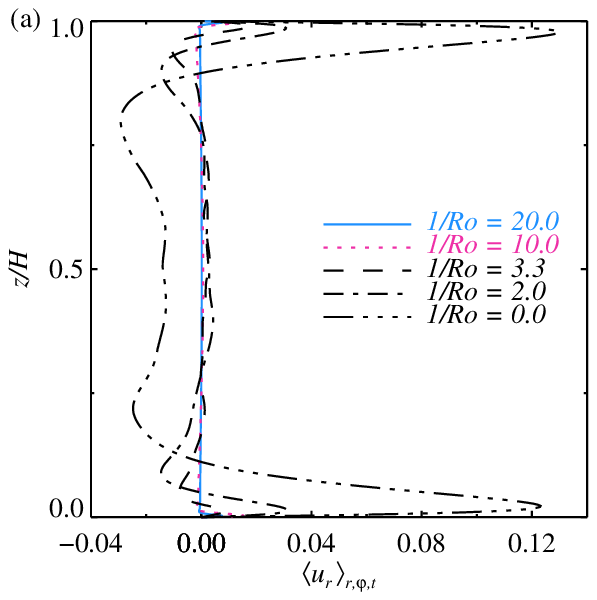} \includegraphics{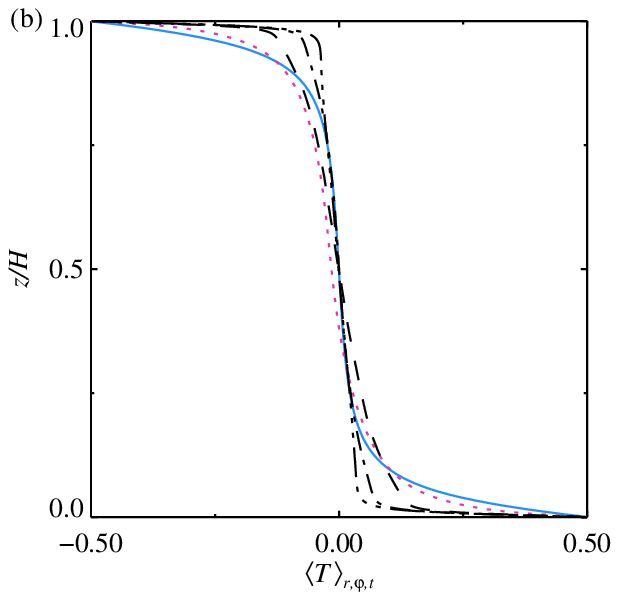}\\
 \includegraphics{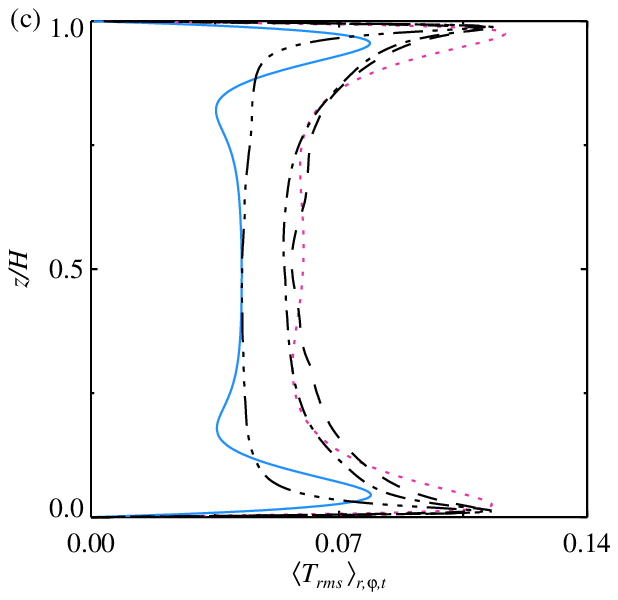}\includegraphics{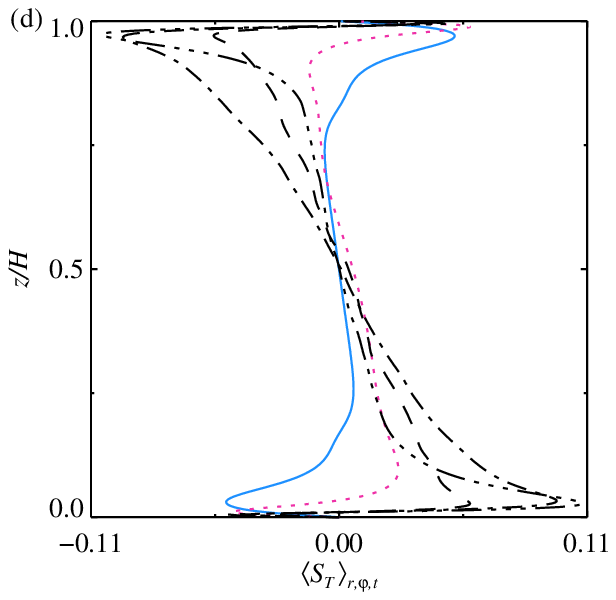} 
\caption{\label{fig:profiles}Temporally, radially and azimuthally averaged profiles of (a) the radial velocity component $u_r$, (b) the temperature $T$, (c) the rms temperature $T_{rms}$ and (d) the skewness of the temperature $S_T$ for $\Ra = 10^8$. The radial averaging was performed for $0 \leq r \leq 0.9R$. In all figures $1/\Ro = 20.0$ is indicated by a blue solid line, $1/\Ro = 10.0$ by a dotted pink line, $1/\Ro = 3.3$ by a short-dashed black line, $1/\Ro = 2.0$ by a dash-dotted black line and $1/\Ro = 0.0$ by dash-triple-dotted black line.}
\end{figure}
That a transition in the flow occurs is also visible in other important flow characteristics  as the radial velocity component $u_r$, the temperature $T$, the rms temperature $T_{\rms}$ or the skewness of the temperature $S_T$. The temporally, radially and azimuthally averaged profiles of these quantities are presented in figure~\ref{fig:profiles}  for five representative rotation rates, $1/\Ro \in \{0.0,2.0,3.3,10.0,20.0\}$. To rule out the sidewall effects \citep{Kunnen2013}, the radial averaging was performed for $0 \leq r \leq 0.9R$ \citep{Stevens2010a}. All of them, but in particular the radial velocity and the skewness of the temperature, reveal a significant flow change for $1/\Ro \gtrsim 3.3$.\par
The radial velocity $u_r$ as function of the vertical coordinate $z$, figure~\ref{fig:profiles}(a), nicely demonstrates the Taylor--Proudman effect. Variations of the flow in vertical direction are inhibited and as a consequence roll-like structures such as the LSC are permitted. Hence, while the $u_r$ mean profiles at low rotation rates still show the typical shape reflecting these structures, they show no variation in the bulk any more as soon as these structures break down. The first three moments of the temperature, presented in figure~\ref{fig:profiles}(b)--(d), reflect the impact of the generated columnar vortices.  The mean temperature profiles exhibit a non-zero gradient in the bulk increasing with $1/\Ro$ and strongest close to the plates. It is usually attributed to vortex merger \cite[]{Julien1996b}. However, even without rotation a small non-vanishing temperature gradient is present that we assume to be due to the small aspect ratio. The rms temperature also varies significantly with the rotation rate. It is almost constant without rotation, showing a crescent-shaped profile up to $1/\Ro \approx 10.0$. For $1/\Ro = 20.0$, the crescent-shape is dented in the midplane and bent in the opposite direction. The skewness of the temperature also  exhibits signs of a fundamental change in the flow, in particular close to the vicinity of the top and bottom plates: with increasing $1/\Ro$, $S_T$ abruptly changes sign at $1/\Ro \approx 10.0$. A similar change of behaviour was also reported by \cite{Kunnen2006,Kunnen2009} for the skewness of the vertical velocity $S_{u_z}$ and of the rms vorticity $S_{\omega_z}$, obtained by simulations in a periodic domain, $\Pran = 1$ and $\Ra = 2.5 \times 10^6$.
\subsection{Toroidal and poloidal potential and energy}
\begin{figure}
\centering
 \begin{overpic}{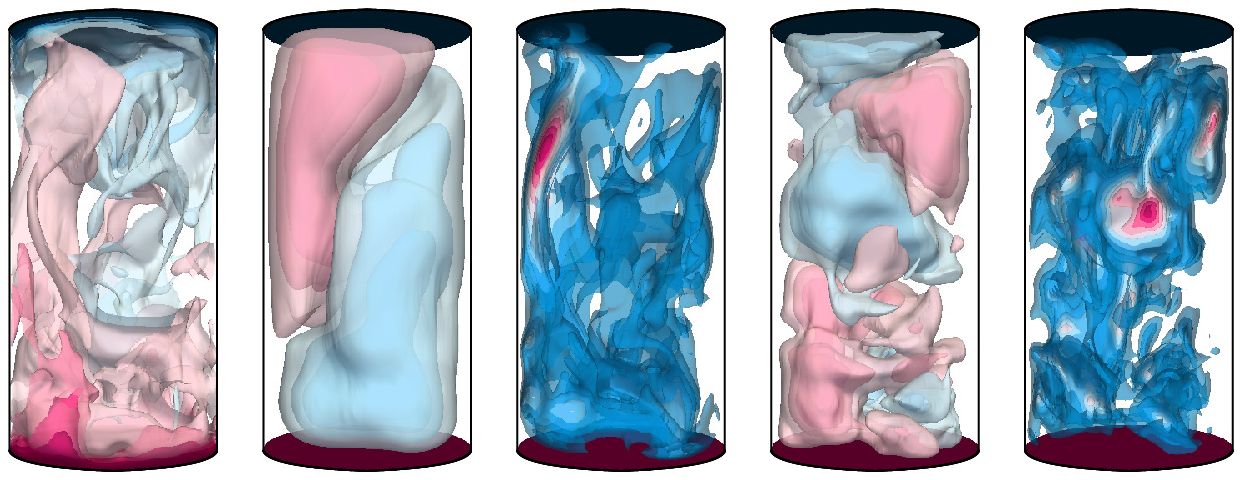}
\put(6,0){$(a)\; T$}
\put(27,0){$(b)\; \xi$}
\put(46,0){$(c)\; e_{\pol}$}
\put(67,0){$(d)\; \psi$}
\put(86,0){$(e)\; e_{\tor}$}
\end{overpic}
\caption{(Colour online) Instantaneous flow structures for $\Ra = 10^8$ and $1/\Ro = 0$, i.e. without rotation. Shown are twelve isosurfaces that are equidistantly distributed between the interval boundaries. Colour scale ranges from blue (the smallest value) through white to pink (the largest value). $(a)$ Temperature $T \in [-0.5,0.5]$, $(b)$ poloidal potential $\xi \in [-0.04,0.04]$, $(c)$ poloidal energy $e_{\pol} \in [0,0.76]$, $(d)$ toroidal potential $\psi \in [-0.13,0.13]$, $(e)$ toroidal energy $e_{\tor} \in [0,0.27]$. \label{fig:3droinf}}
\end{figure}
\begin{figure}
\centering
 \begin{overpic}{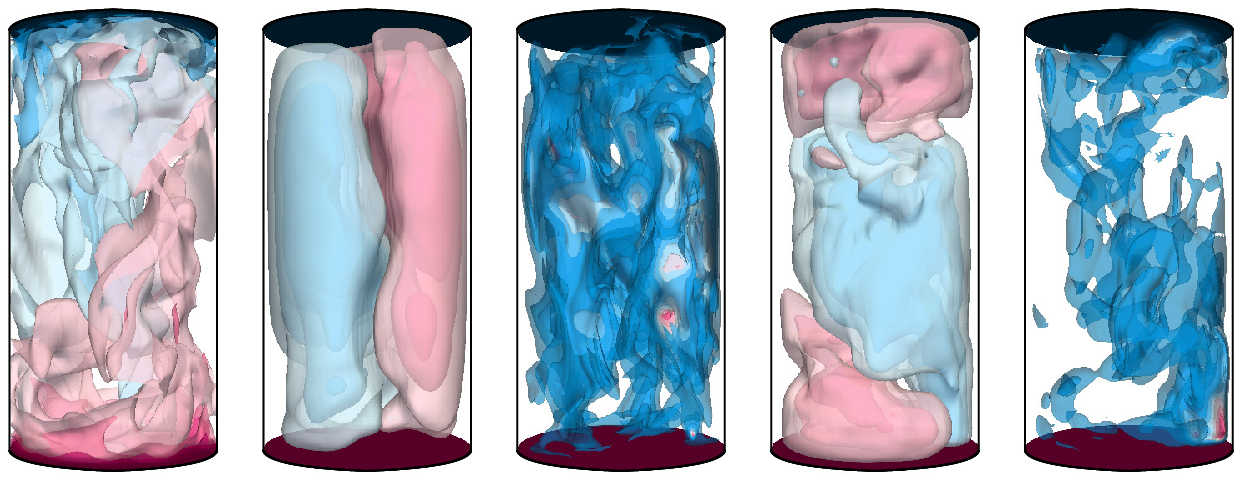}
\put(6,0){$(a)\; T$}
\put(27,0){$(b)\; \xi$}
\put(46,0){$(c)\; e_{\pol}$}
\put(67,0){$(d)\; \psi$}
\put(86,0){$(e)\; e_{\tor}$}
\end{overpic}
\caption{(Colour online) As in figure~\ref{fig:3droinf}, but for $1/\Ro = 2.0$. $(a)$ Temperature $T \in [-0.5,0.5]$, $(b)$ poloidal potential $\xi \in [-0.03,0.03]$, $(c)$ poloidal energy $e_{\pol} \in [0,0.70]$, $(d)$ toroidal potential $\psi \in [-0.16,0.16]$, $(e)$ toroidal energy $e_{\tor} \in [0,0.74]$. \label{fig:3dro05}}
\end{figure}
\begin{figure}
\centering
 \begin{overpic}{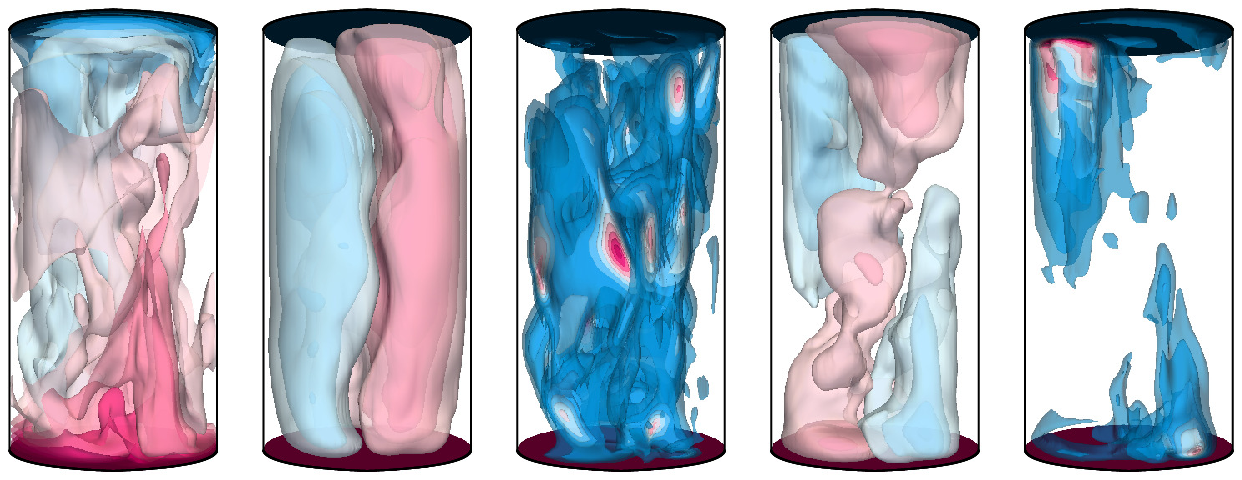}
\put(6,0){$(a)\; T$}
\put(27,0){$(b)\; \xi$}
\put(46,0){$(c)\; e_{\pol}$}
\put(67,0){$(d)\; \psi$}
\put(86,0){$(e)\; e_{\tor}$}
\end{overpic}
\caption{(Colour online) As in figure~\ref{fig:3droinf}, but for $1/\Ro = 3.3$. $(a)$ Temperature $T \in [-0.5,0.5]$, $(b)$ poloidal potential $\xi \in [-0.02,0.02]$, $(c)$ poloidal energy $e_{\pol} \in [0,0.27]$, $(d)$ toroidal potential $\psi \in [-0.21,0.21]$, $(e)$ toroidal energy $e_{\tor} \in [0,0.77]$. \label{fig:3dro03}}
\end{figure}
\begin{figure}
\centering
 \begin{overpic}{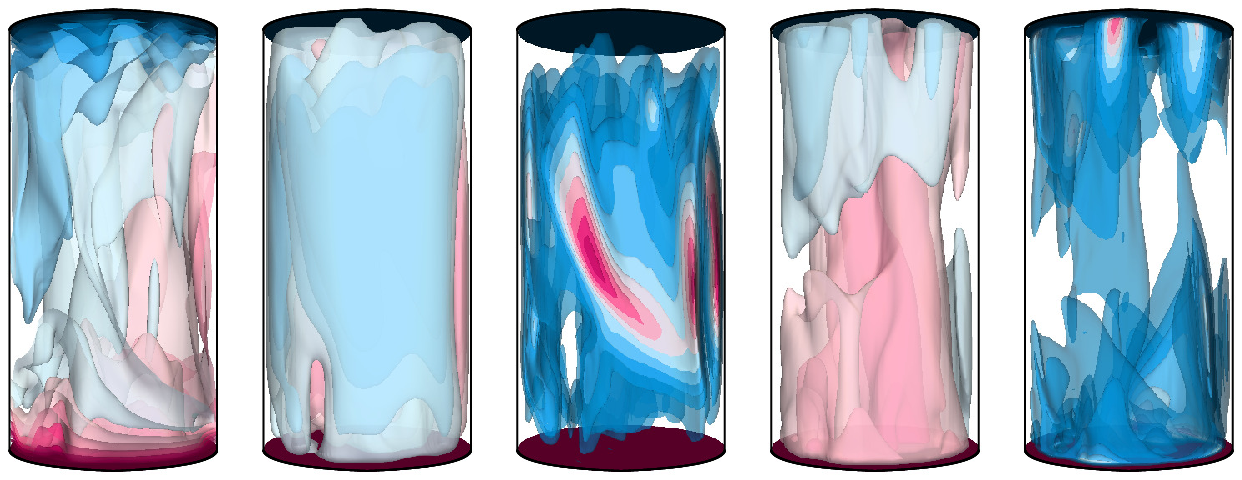}
\put(6,0){$(a)\; T$}
\put(27,0){$(b)\; \xi$}
\put(46,0){$(c)\; e_{\pol}$}
\put(67,0){$(d)\; \psi$}
\put(86,0){$(e)\; e_{\tor}$}
\end{overpic}
\caption{(Colour online) As in figure~\ref{fig:3droinf}, but for $1/\Ro = 10.0$.  $(a)$ Temperature $T \in [-0.5,0.5]$, $(b)$ poloidal potential $\xi \in [-0.003,0.003]$, $(c)$ poloidal energy $e_{\pol} \in [0,0.12]$, $(d)$ toroidal potential $\psi \in [-0.11,0.11]$, $(e)$ toroidal energy $e_{\tor} \in [0,0.22]$.\label{fig:3dro01} }
\end{figure}
\begin{figure}
\centering
 \begin{overpic}{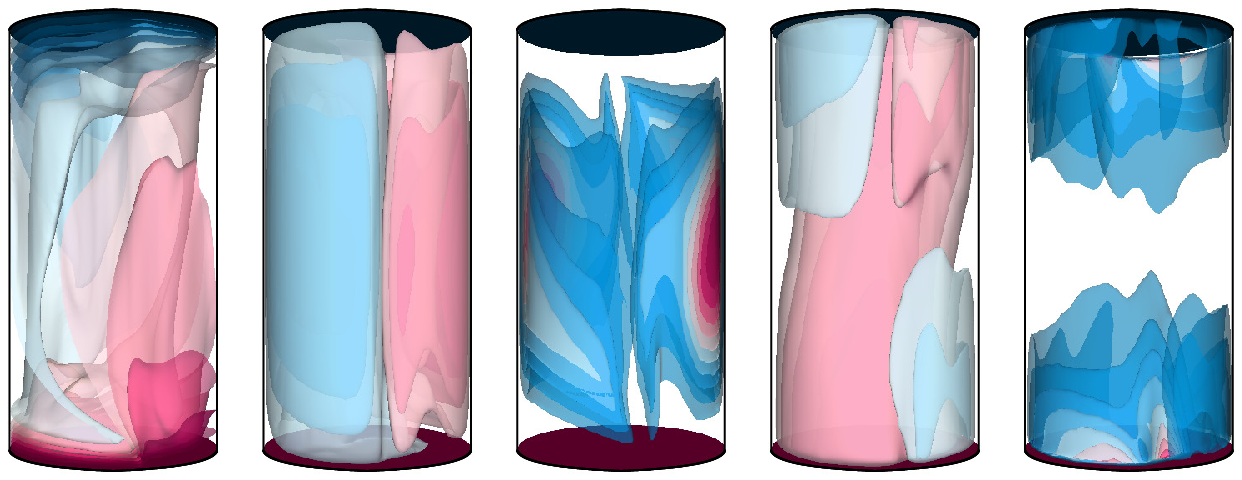}
\put(6,0){$(a)\; T$}
\put(27,0){$(b)\; \xi$}
\put(46,0){$(c)\; e_{\pol}$}
\put(67,0){$(d)\; \psi$}
\put(86,0){$(e)\; e_{\tor}$}
\end{overpic}
\caption{(Colour online) As in figure~\ref{fig:3droinf}, but for $1/\Ro = 20.0$. $(a)$ Temperature $T \in [-0.5,0.5]$, $(b)$ poloidal potential $\xi \in [-0.001,0.001]$, $(c)$ poloidal energy $e_{\pol} \in [0,0.05]$, $(d)$ toroidal potential $\psi \in [-0.03,0.03]$, $(e)$ toroidal energy $e_{\tor} \in [0,0.09]$. \label{fig:3dro005} }
\end{figure}
 In the following we show how to connect the different flow behaviour with the toroidal and poloidal potential and energy. It will allow to effectively identify the transitions between different regimes in rotating Rayleigh--B\'enard based on the global quantities $e_{\tor}$ and $e_{\pol}$.\par
In the figures~\ref{fig:3droinf}--\ref{fig:3dro005} we present instantaneous flow quantities for the same representative rotation rates $1/\Ro \in \{0.0,2.0,3.3,10.0,20.0\}$ at $\Ra = 10^8$ as before. We refrained from showing averaged flow fields, since the precession motion of the flow might distort their interpretation.\par
In the non-rotating case,  $1/\Ro = 0.0$, the flow is most of the time organised in an LSC. This structure is not only visible in the temperature field $T$ but also in the poloidal potential $\xi$ and energy $e_{\pol}$. The toroidal potential $\psi$ shows a rather chaotic structure and the toroidal energy $e_{\tor}$ is concentrated in the bulk and of lower magnitude than $e_{\pol}$. If the flow was two-dimensional and independent from one horizontal direction then it would be completely poloidal. However, since the toroidal energy is associated with the vortices in the flow, the toroidal field acts in a destabilising way on the flow and tears the plumes apart. On the other hand, when the convection cell is rotated, the toroidal field has a stabilising effect. With increasing rotation rate, as seen in the figures~\ref{fig:3dro05}--\ref{fig:3dro005}, the flow resembles a horizontally two-dimensional flow. But a two-dimensional flow independent from the vertical direction is fully toroidal. This means that with increasing $1/\Ro$ the toroidal energy increases and the poloidal energy decreases. Instead of an LSC, there are elongated flow structures emerging from the edge of the boundary layers, that are visible in the temperature field and in the toroidal potential $\psi$ and energy $e_{\tor}$. The highest toroidal energy is contained in these short columnar-like vortices seen in figure~\ref{fig:3dro03} and \ref{fig:3dro01}. For even more rapid rotation, at $1/\Ro = 20.0$, which is close to the onset of convection, wall modes dominate and both the poloidal and toroidal energy are highest close to the sidewall. Hence, the toroidal and poloidal energy can be used to characterize the different types of dynamics in rotating Rayleigh--B\'enard convection.\par
\begin{figure}
 \includegraphics{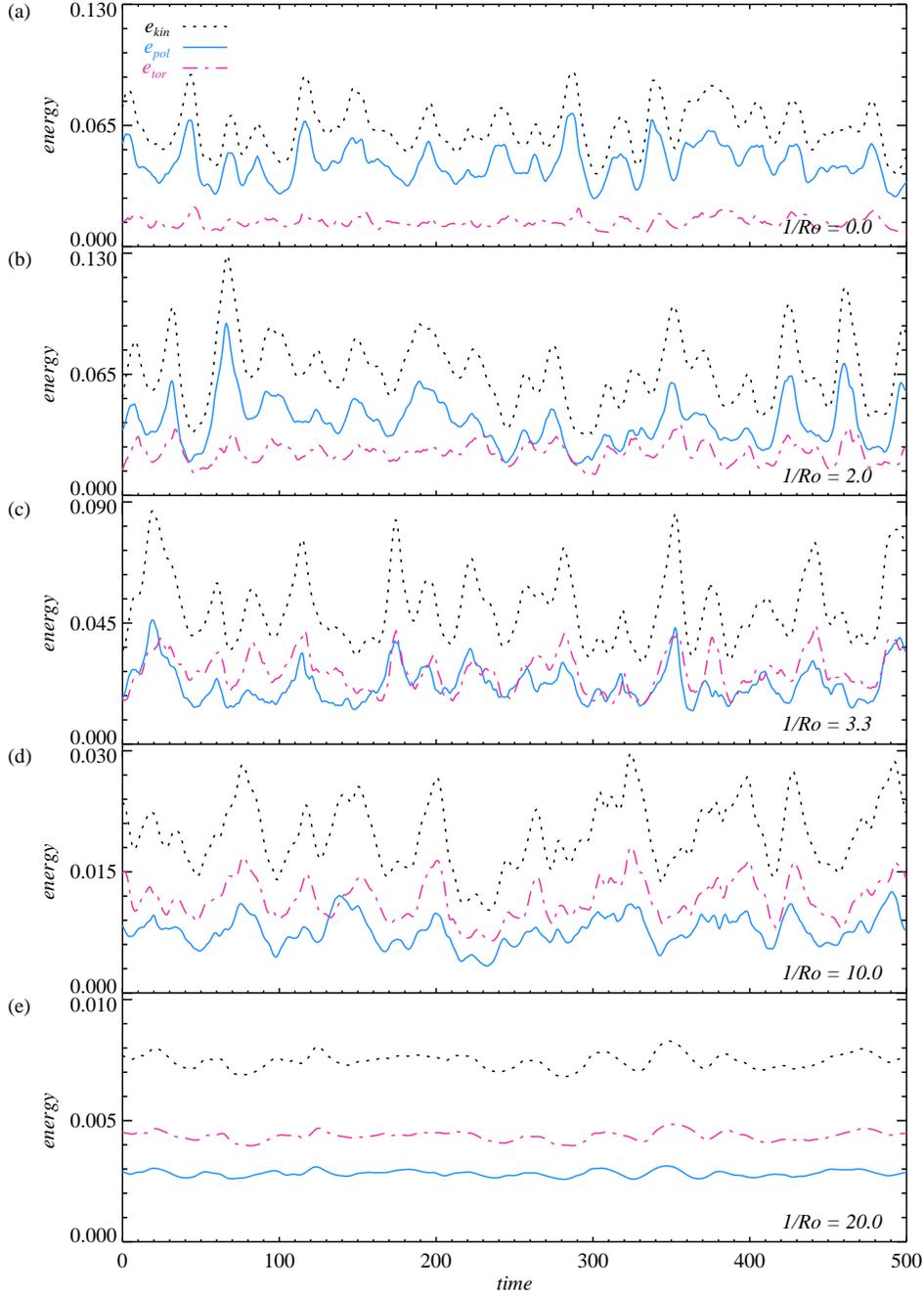}
\caption{\label{fig:time}(Colour online) Times series of the volume-averaged kinetic energy $e_{\kin}$ (black dotted line), the poloidal energy $e_{\pol}$ (blue solid line) and the toroidal energy $e_{\tor}$ (pink dash-dotted line) for $\Ra = 10^8$ and (a)~$1/\Ro = 0.0$, (b)~$1/\Ro = 2.0$, (c)~$1/\Ro = 3.3$, (d)~$1/\Ro = 10.0$, (e)~$1/\Ro = 20.0$. The time is measured in dimensionless time-units (see main text), and all time series were deliberately set to zero at a point when statistical equilibrium was reached.}
\end{figure}
In figure \ref{fig:time} we also show the temporal evolution of $e_{\kin}$, $e_{\pol}$ and $e_{\tor}$. Indeed for $1/\Ro = 0.0$ and $1/\Ro = 2.0$ holds $e_{\pol} > e_{\tor}$  for all instances of time, while for $1/\Ro = 10.0$ and $1/\Ro = 20.0$ it is $e_{\pol} < e_{\tor}$.  For $1/\Ro = 3.3$, $e_{\pol}$ and $e_{\tor}$ are of the same order. Furthermore, the oscillation frequency tends to decrease together with $\Ro$.\par
To analyse these observations quantitatively, the volume- and time-averaged energies are shown in figure \ref{fig:vol}  for all conducted simulations. Similar as in figure~\ref{fig:Nu}~(a), we also compare them with $1/\Ro_b$ suggested by \citet{Weiss2010} and \citet{Weiss2011a} and the empirically found $1/\Ro_1$ and $1/\Ro_2$ by \citet{Ecke2013}.\par
The DNS data share some common features for all Rayleigh numbers. All energies $e_{\kin}$, $e_{\pol}$ and $e_{\tor}$ are independent of $1/\Ro$ up to approximately $1/\Ro_b = 0.86$. At this point, $e_{\kin}$ and $e_{\pol}$ decrease monotonically. On the contrary, $e_{\tor}$ increases at this point, then reaches a maximum and after that drops with increasing $1/\Ro$. It reaches the same value as in the non-rotating case at $1/\Ro_2 \approx 8.33$.\par
For the relatively low Rayleigh numbers $\Ra = 10^5$ and $\Ra = 10^6$, the toroidal energy is always lower than the poloidal energy, despite the fact that it can be up to about eight and three times higher, respectively, compared to the non-rotating case as it is recognizable in figure~\ref{fig:vol}(a)--(d). As a consequence, the poloidal field is for all rotation rate able to sustain cellular-like flow structures. Since for $\Ra = 10^5$ convection is steady for all $1/\Ro$, the standard deviation $\sigma$ is zero. For $\Ra = 10^6$, convection is unsteady for $0 \leq 1/\Ro \lesssim 1.67$ and due to the small computational mesh, DNS for several thousands of time units could be performed. Hence, the large error bars in figure~\ref{fig:vol}(b) and (d) indicate physical variations and not a lack of statistics. In the range $1.67 \lesssim 1/\Ro \lesssim 2.5$, when the convective heat transport is oscillatory, $\sigma$ decreases with increasing $1/\Ro$. The oscillatory behaviour is naturally present in the time series of $e_{\kin}$, $e_{\pol}$ and $e_{\tor}$. For even larger $1/\Ro$, we found steady convection, in a sense that the Nusselt number does not change in time. Interestingly, at the transition between oscillatory and steady convection, there is a minimum in the toroidal energy and this point also coincides with $1/\Ro_1$.\par
For $\Ra = 10^7$ we observed turbulent convection for all considered $1/\Ro$, except for the highest rotation rate $1/\Ro = 14.1$, where convection is steady. But in this case, presented in \ref{fig:vol}(e) and (f), the toroidal energy is higher than the poloidal energy for $1/\Ro$ being greater than a critical inverse Rossby number $1/\Ro_{\crit}$. This critical inverse Rossby number is hence determined by the condition
\begin{equation}
\frac{1}{\Ro_{\crit}} = \left.\frac{1}{\Ro}\right|_{e_{\pol} = e_{\tor}}.
\end{equation}
We argue, that only if toroidal motions are prevailing, i.e. $e_{\tor} > e_{\pol}$, one can speak of rotation dominated convection. If, on the other side, poloidal motions, i.e. $e_{\tor} < e_{\pol}$, are predominant then buoyancy is more important. If  $e_{\tor} > e_{\pol}$, the LSC or other roll-like structures cease to exist and instead columnar vortices become apparent. Thus, this clarifies the change of behaviour, observed in the global flow properties presented in figure \ref{fig:profiles}  and discussed in the previous section. Furthermore, these findings are also in agreement with those by other authors \citep{Stevens2012, Stevens2013a, Kunnen2008a} that relate the breakdown of the large-scale circulation to the regime of rotation dominance. They also shed some more light on the fact, why not only the inverse Rossby number but also the Rayleigh number has to be sufficiently high to be in a rotation dominated regime \citep{Julien2012, Ecke2013}. The behaviour of $e_{\kin}$, $e_{\pol}$ and $e_{\tor}$ with $1/\Ro$ at $\Ra = 10^8$ and $\Ra = 10^9$, displayed in figure~\ref{fig:vol}(g)--(j), is very similar to that of $\Ra = 10^7$, but the maximum relative enhancement of the toroidal energy compared to the non-rotating case is diminished with higher $\Ra$. Nonetheless, in these cases the crossover of the poloidal and toroidal energy is more pronounced, in a sense that the difference between $e_{\tor}$ and $e_{\pol}$ is larger at rapid rotation.\par
To determine the transition point more accurately, the ratio of the toroidal to the total kinetic energy $e_{\tor}/e_{\kin}$ and the ratio of the poloidal to the total kinetic energy $e_{\pol}/e_{\kin}$ are shown in figure \ref{fig:ratio}. Besides, the ratios $e_{\tor}/e_{\kin}$ and $e_{\pol}/e_{\kin}$ are known to be properties of the flow characterizing the different types of dynamics in non-rotating Rayleigh--B\'enard convection \cite[]{Breuer2004} and we argue that the same is true for rotating Rayleigh--B\'enard convection.\par
Figure~\ref{fig:ratio} reveals various information. First of all, the critical inverse Rossby number is about $1/\Ro_\crit \approx 3.0$ or more accurately it lies in the range $2.36 \lesssim 1/\Ro_{\crit}  \lesssim 3.33$. At this point, both the poloidal and the toroidal energy are about 50\% of the total kinetic energy. From the equations \eqref{eq:dec1}--\eqref{eq:dec3} and \eqref{eq:ekin}--\eqref{eq:etor} it is obvious that not all of the kinetic energy is contained in the toroidal and poloidal part, since
\begin{equation}
 e_\kin \neq e_\pol + e_\tor.
\end{equation}
\begin{figure}
\centering
\includegraphics{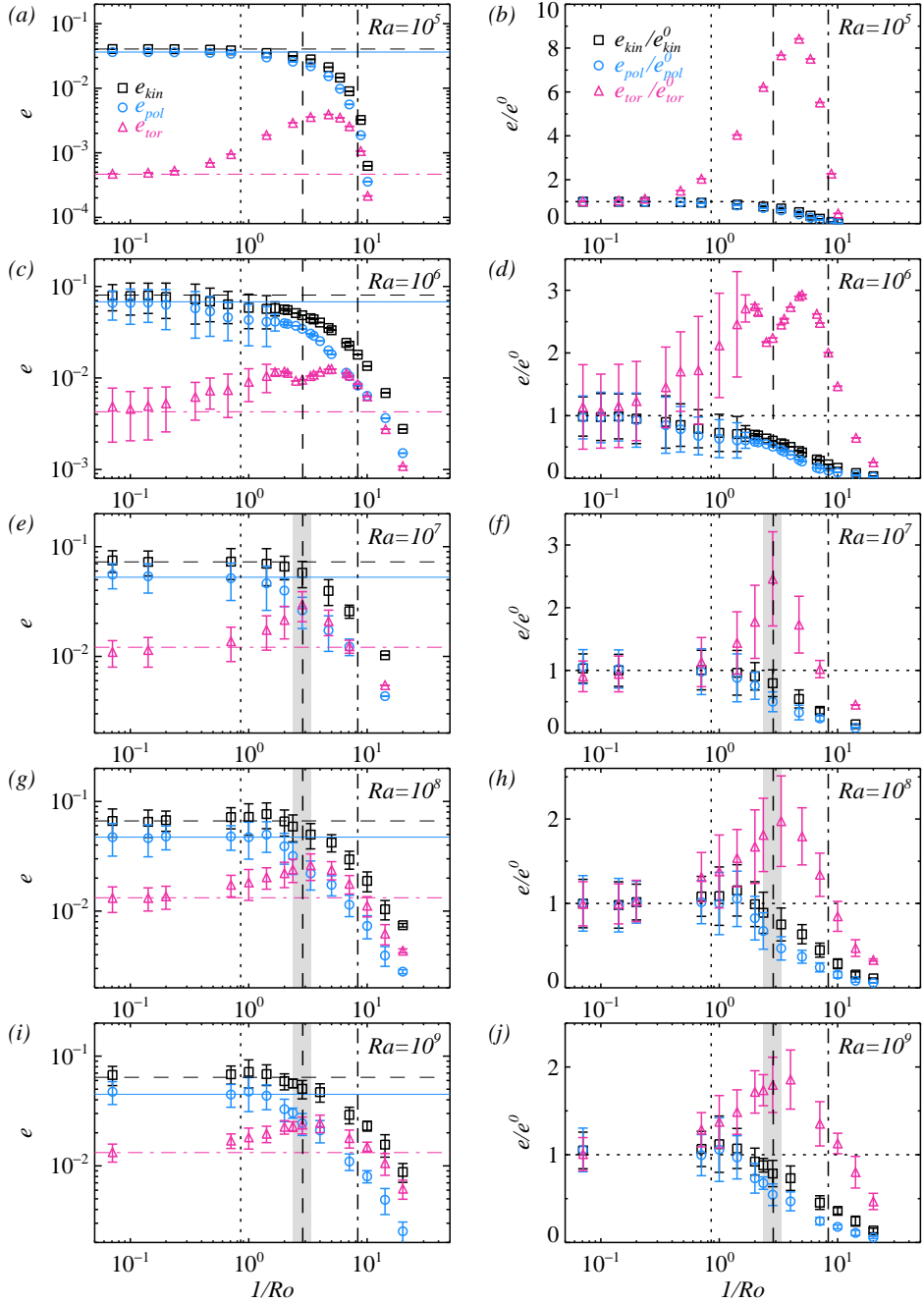}
\caption{\label{fig:vol}(Colour online) Left panel: Volume- and time-averaged kinetic energy $e_{\kin}$ (black squares and dashed line), poloidal energy $e_{\pol}$ (blue circles and solid line) and toroidal energy $e_{\tor}$ (pink triangles and dash-dotted line) as function of $1/\Ro$. The horizontal lines indicate the value in the non-rotating case. Right panel: Volume- and time-averaged kinetic energies normalised by their value in the non-rotating case. The vertical dotted line shows prediction by \citet{Weiss2010,Weiss2011a}, $1/\Ro_b = 0.86$, the vertical long-dashed and dashed-dotted line mark the proposed transition by \cite{Ecke2013} at $1/\Ro_1 \approx 2.86$ and $1/\Ro_2 \approx 8.33$, respectively. The grey shaded area indicates where $e_{\pol} \approx e_{\tor}$ at $2.36 \lesssim 1/\Ro_{\crit}  \lesssim 3.33$. The error bars show the standard deviation $\sigma$ of the averaged values.}
\end{figure}%
However, as can be readily seen from figure~\ref{fig:ratio}, the sum of $e_\tor$ and $e_\pol$ is for all cases about 90\% of the total kinetic energy, indicating that the single components of the toroidal and poloidal field are almost uncorrelated.\par
At small inverse Rossby numbers, $1/\Ro \lesssim 1$ and also in non-rotating convection, the poloidal energy decreases, and the toroidal energy increases with the Rayleigh number. Without rotation, at $\Ra = 10^5$ about 90\% of the kinetic energy is contained in the poloidal energy and only about 1\% in the toroidal energy. At $\Ra = 10^9$ only about 70\% of the kinetic energy is contained in the poloidal motion and 20\% in the toroidal energy. This is not surprising, because the higher $\Ra$ the higher the  number of plumes. Hence, there is an increased shearing and swirling in the flow that is associated with a vertical vorticity and a higher $e_\tor$. Consequently, $e_\pol$ has to decrease. This is also related to the picture of a less strong LSC at higher $\Ra$. With increasing $1/\Ro$ one has to distinguish between the steady cases $\Ra = 10^5$ and $10^6$ and the turbulent cases $10^7 \leq \Ra \leq 10^9$.  At intermediate inverse Rossby numbers, $1 \lesssim 1/\Ro \lesssim 5$, $e_\tor/e_\kin$ and $e_\pol/e_\kin$ collapse at approximately the same value for $10^7 \leq \Ra \leq 10^9$, unlike for $10^5$ and $10^6$. Between $7 \lesssim 1/\Ro \lesssim 10$ the two data sets of $\Ra = 10^6$ and $10^7$ cling to each other, however, for $\Ra = 10^6$, $e_{\tor}/e_{\kin}$ drops afterwards to a value of about 0.4 at $1/\Ro = 20.0$ and $e_{\pol}/e_{\kin}$ raises again to a value of about 0.55. On the contrary, $e_{\tor}/e_{\kin}$ increases further to 0.53 and $e_{\pol}/e_{\kin}$ decreases to 0.42 at $1/\Ro = 14.1$ for $\Ra = 10^7$. Thus, there is a clear distinction between non-turbulent and turbulent rotating Rayleigh--B\'enard convection. Like this, at large inverse Rossby numbers, $1/\Ro \gtrsim 5$, the data show a larger spread depending on the Rayleigh number. At $1/\Ro \gtrsim 14$, $\Ra =10^5$ is in the conducting state, hence $e_\tor$, $e_\pol$ and $e_\kin$ are zero. The relative toroidal energy is highest for $\Ra = 10^9$, being about 70\% of the kinetic energy and the poloidal energy is lowest for the very same $\Ra$, being about 29\% at $1/\Ro = 20.0$.\par
Another way of collapsing the data has been suggested by \cite{Ecke2013} choosing the quantity $\Ra \Ek^{7/4} = \Ra^{1/8} \Pran^{7/8} \Ro^{7/4}$ instead of $1/\Ro$. This implies a dependence on $\Ra$ and $\Pran$, and indeed figure~\ref{fig:ecke}~(a) reveals that the collapse of the critical point where the toroidal and poloidal energy are equal is even better. The in this way determined crossover happens in the range $1 \leq \Ra \Ek^{7/4} \leq 2$ or at $\Ra \Ek^{7/4} \approx 1.5$, respectively. This quantity was also found to be a suitable scaling variable in water with $\Pran = 7$ \citep{King2009}, although it was corrected to $\Ra \Ek^{3/2}$ later on \citep{King2012}. However, the latter does not fit to our data. A better agreement of the $\Nu$ behaviour for all $\Ra$ (except for the particularities occurring at $\Ra = 10^6$) when plotted against $\Ra \Ek^{7/4}$ is also true and had already been found by \cite{Ecke2013}. It is presented in figure~\ref{fig:ecke}~(b).
\begin{figure}
\includegraphics{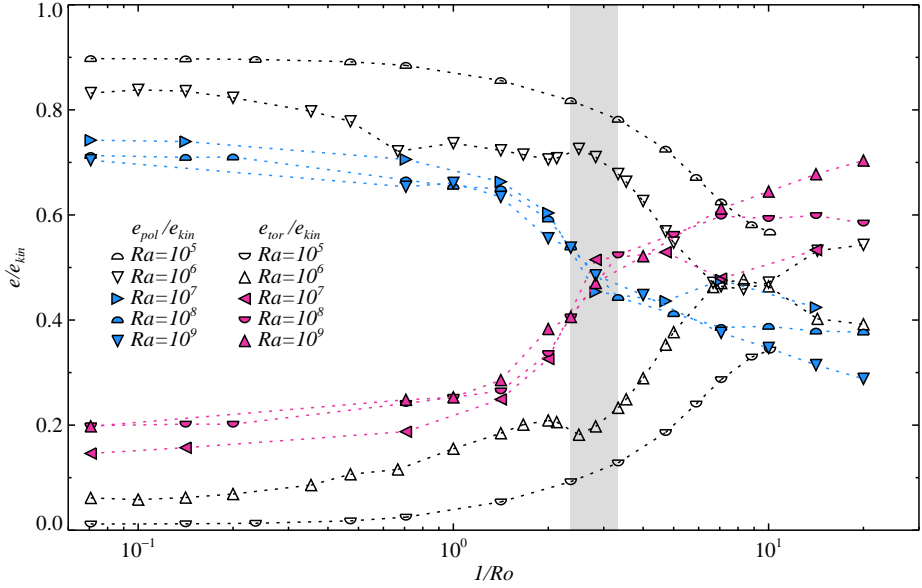}
\caption{\label{fig:ratio}(Colour online) Poloidal and toroidal energy as fraction of the total kinetic energy versus $1/\Ro$. The grey shaded area shows the approximate range where $e_{\pol} = e_{\tor}$ at $2.4 \lesssim 1/\Ro_{\crit}  \lesssim 3.3$.}
\end{figure}%
\begin{figure}
\begin{overpic}{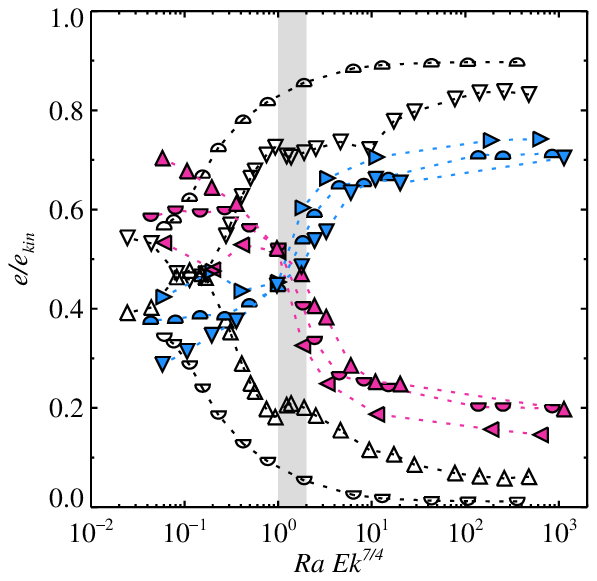}
 \put(0,92){(a)}
 \end{overpic}
 \hfill
\begin{overpic}{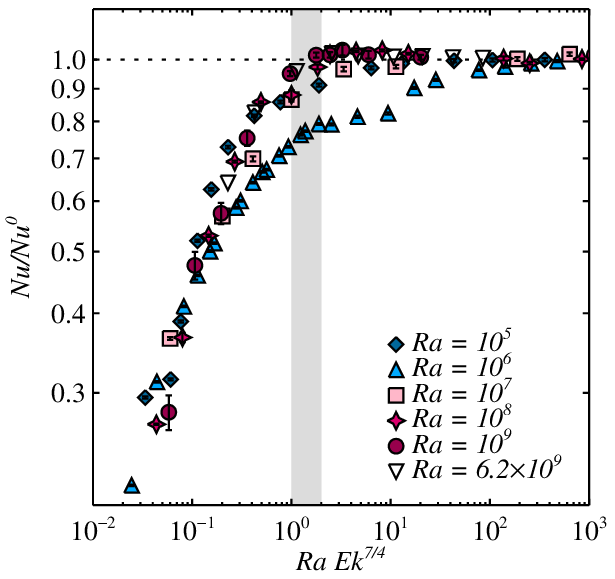}
  \put(0,92){(b)}
\end{overpic}
\caption{\label{fig:ecke}(Colour online) (a) Poloidal and toroidal energy as fraction of the total kinetic energy versus $\Ra \Ek^{7/4}$. Similar as figure~\ref{fig:ratio}. The grey shaded area shows the approximate range where $e_{\pol} = e_{\tor}$ at $1.0 \lesssim \Ra \Ek_\crit^{7/4} \lesssim 2.0$. (b) Nusselt number for the rotating case normalized by the one in the non-rotating case $\Nu/\Nu^0$ as function of $\Ra \Ek^{7/4}$. Similar as figure~\ref{fig:Nu}~(a).}
\end{figure}%
\section{Concluding remarks}
Rotating Rayleigh--B\'enard convection of a fluid with a Prandtl number of $\Pran = 0.8$ in a slender cylinder with an aspect ratio of $\Gamma = 0.5$ was studied in the Rayleigh number range $10^5 \leq \Ra \leq  10^9$.   The rotation rate was varied between the inverse Rossby numbers $0$ and $20$. Depending on the rotation rate, the general flow phenomenology changes, and with it certain flow characteristics, such as the temperature, the radial velocity, the rms temperature and the skewness of the temperature.  These changes are not clearly present in the behaviour of the Nusselt number $\Nu$  even though also there different scalings depending on $1/\Ro$ are observed \citep{Ecke2013}. To identify these regime transitions, we decomposed the velocity field into its toroidal and poloidal scalar field, and analysed the contribution of $e_{\tor}$ and $e_{\pol}$ to the total kinetic energy $e_{\kin}$. Evaluating regime transitions by means of $e_{\tor}$ and $e_{\pol}$ has the advantage that it bases on global quantities which are characteristic for the flow in rotating and in non-rotating turbulent thermal convection. The poloidal energy is associated with all cellular-like structures, such as the LSC or multiple roll state, i.e. the flow typically observed in non-rotating convection. The toroidal energy is associated with the vertical vorticity and hence with columnar vortices, typical in rotating convection. Hence, this method is expected to work independently of the aspect ratio and of the Prandtl number and forthcoming studies with different $\Gamma$ and $\Pran$ are to be conducted, to reinforce this idea.\par
In the present DNS we can identify four different regimes with the proposed method. As long as $e_{\tor}$ has the same value as in the non-rotating case, i.e. $e_{\tor}/e^0_\tor = 1$, the flow is completely dominated by buoyancy. As soon as the toroidal energy relative to the non-rotating case increases, $e_{\tor}/e^0_\tor > 1$, Rayleigh--B\'enard convection is considered to be rotation influenced. This agrees well with the bifurcation point found by \cite{Weiss2010} and \cite{Weiss2011a}, which gives $1/\Ro_b = 0.86$ for a cylindrical $\Gamma = 0.5$ cell. At a rotation rate where the toroidal energy is greater than the poloidal one, convection is rotation dominated and large-scale roll structures, such as the LSC, are expected to cease to exist and instead columnar vortex structures dominate the flow. To reach this regime, however, the Rayleigh number has to be at least about $10^7$. The critical inverse Rossby number is thus determined by the condition $e_{\pol} = e_{\tor}$ which gives $1/\Ro_{\crit} = \left.1/\Ro\right|_{e_{\pol} = e_{\tor}} \approx 3$ ($2.36 \lesssim 1/\Ro_{\crit}  \lesssim 3.33$) for the cases considered. Finally, when the toroidal energy drops below the value of the non-rotating case, $e_{\tor}/e^0_\tor < 1$ one reaches the regime of geostrophic turbulence. The last two in this way determined transitions agree well with the ones found by \cite{Ecke2013}, who identify them by a different scaling behaviour of $\Nu$, finding $1/\Ro_1 \approx 2.86$ and $1/\Ro_2 \approx 8.33$. An even better collapse of data for the Nusselt number and the toroidal and poloidal energy for all $\Ra$ considered can be obtained by using $\Ra \Ek^{7/4}$ instead of $1/\Ro$ as scaling variable, yielding 
$1.0 \lesssim \Ra \Ek_\crit^{7/4} \lesssim 2.0$ for the critical rotation rate where $e_\pol \approx e_\tor$.

\section*{Acknowledgement}
SH wishes to thank Keith Weinman for fruitful discussions about the numerical solution of Poisson equations. The authors acknowledge the support by the Deutsche Forschungsgemeinschaft (DFG) in the framework of SFB 963/1, project A6, and Heisenberg fellowship SH405/4.

\bibliographystyle{jfm}
\bibliography{rotation}

\end{document}